\documentstyle[psfig,psfrag,epsf,epsfig]{siamltex}

\title{Full Spectrum of Lyapunov Exponents in Gauge Field Theory}

\author{Tam\'as S. Bir\'o\thanks{Research Institute for
        Particle and Nuclear Physics, Hungarian Academy of Sciences, H-1525 Budapest, Hungary}
        \and Harald Markum\thanks{Atominstitut, Technische Universit\"at Wien, A-1040 Vienna, Austria}
        \and Rainer Pullirsch\thanks{Atominstitut, Technische Universit\"at Wien, A-1040 Vienna, Austria}}

\begin{document}

\maketitle

\begin{abstract} 
        We analyze the Lyapunov exponents of U(1) gauge fields across the
        phase transition from the confinement to the Coulomb phase on the
        lattice which are initialized by quantum Monte Carlo simulations.
        We observe all features of a strange attractor with a tendency to
        regularity towards the continuum limit. Results are also
        displayed for the full spectrum of Lyapunov exponents of the SU(2)
        gauge system.
\end{abstract}

\pagestyle{myheadings}
\thispagestyle{plain}
\markboth{T.S Bir\'o, H. Markum, and R. Pullirsch}{Full Spectrum of Lyapunov
          Exponents in Gauge Field Theory}

\section{Classical chaotic dynamics from quantum Monte Carlo
         initial states}

Chaotic dynamics in general is characterized by the
spectrum of Lyapunov exponents. These exponents, if they are positive,
reflect an exponential divergence of initially adjacent configurations.
In case of symmetries inherent in the Hamiltonian of the system
there are corresponding zero values of these exponents. Finally
negative exponents belong to irrelevant directions in the phase
space: perturbation components in these directions die out
exponentially. Pure random gauge fields on the lattice show a characteristic
Lyapunov spectrum consisting of one third of each kind of
exponents~\cite{BOOK}.

The general definition of the Lyapunov exponent is based on a
distance measure $d(t)$ in phase space,
\begin{equation}
L := \lim_{t\rightarrow\infty} \lim_{d(0)\rightarrow 0}
\frac{1}{t} \ln \frac{d(t)}{d(0)}.
\end{equation}
In case of conservative dynamics the sum of all Lyapunov exponents
is zero according to Liouville's theorem, $\sum L_i = 0$.
For gauge field theories one utilizes the gauge invariant distance
measure consisting of
the local differences of energy densities between two three-dimensional
field configurations on the lattice:
\begin{equation}
d : = \frac{1}{N_P} \sum_P\nolimits \, \left| {\rm tr} U_P - {\rm tr} U'_P \right|.
\end{equation}
Here the symbol $\sum_P$ stands for the sum over all $N_P$ plaquettes,
so this distance is bound in the interval $(0,2N)$ for the group
SU(N). $U_P$ and $U'_P$ are the familiar plaquette variables, constructed from
the basic link variables $U_{x,i}$,
\begin{equation}
U_{x,i} = \exp \left( aA_{x,i}T \right)\: ,
\end{equation}
located on lattice links pointing from the position $x=(x_1,x_2,x_3)$ to
$x+ae_i$. The generator of the group U(1) is
$T = -ig$ and $A_{x,i}$ is the vector potential.
The elementary plaquette variable is constructed for a plaquette with a
corner at $x$ and lying in the $ij$-plane as
$U_{x,ij} = U_{x,i} U_{x+i,j} U^{\dag}_{x+j,i} U^{\dag}_{x,j}$.
It is related to the magnetic field strength $B_{x,k}$:
\begin{equation}
U_{x,ij} = \exp \left( \varepsilon_{ijk} a^2 B_{x,k} T \right).
\end{equation}
The electric field strength $E_{x,i}$ is related to the canonically conjugate
momentum $P_{x,i} = a \dot{U}_{x,i}$ via
\begin{equation}
E_{x,i} = \frac{1}{g^2a^2} \left( T \dot{U}_{x,i} U^{\dag}_{x,i} \right).
\end{equation}

The Hamiltonian of the lattice gauge field system can be casted into
the form
\begin{equation}
H = \sum \left[ \frac{1}{2} \langle P, P \rangle \, + \,
 1 - \frac{1}{4} \langle U, V \rangle \right].
\end{equation}
Here the scalar product stands for
$\langle A, B \rangle = {\rm Re} (A B^{\dag} )$.
The staple variable $V$ is a sum of triple products of elementary
link variables closing a plaquette with the chosen link $U$.
This way the Hamiltonian is formally written as a sum over link
contributions and $V$ plays the role of the classical force
acting on the link variable $U$. 

We prepare the initial field configurations
from a standard four-dimensional Euclidean Monte Carlo program on
an $N^3\times 4$ lattice varying the inverse gauge coupling
$\beta\propto g^{-2}$~\cite{SU2}.
We relate such four-dimensional Euclidean
lattice field configurations to Minkow\-skian momenta and fields
for the three-dimensional Hamiltonian simulation
by selecting a fixed time slice of the four-dimensional lattice.

\section{Spectrum of the stability matrix}

\begin{figure*}[ht]
\centerline{{\hspace*{1mm}{\footnotesize U(1) complex Lyapunov spectrum \hspace*{10mm} U(1) complex Lyapunov spectrum}}}\vspace*{-13.5mm}
\centerline{{\hspace*{13mm}\psfig{figure=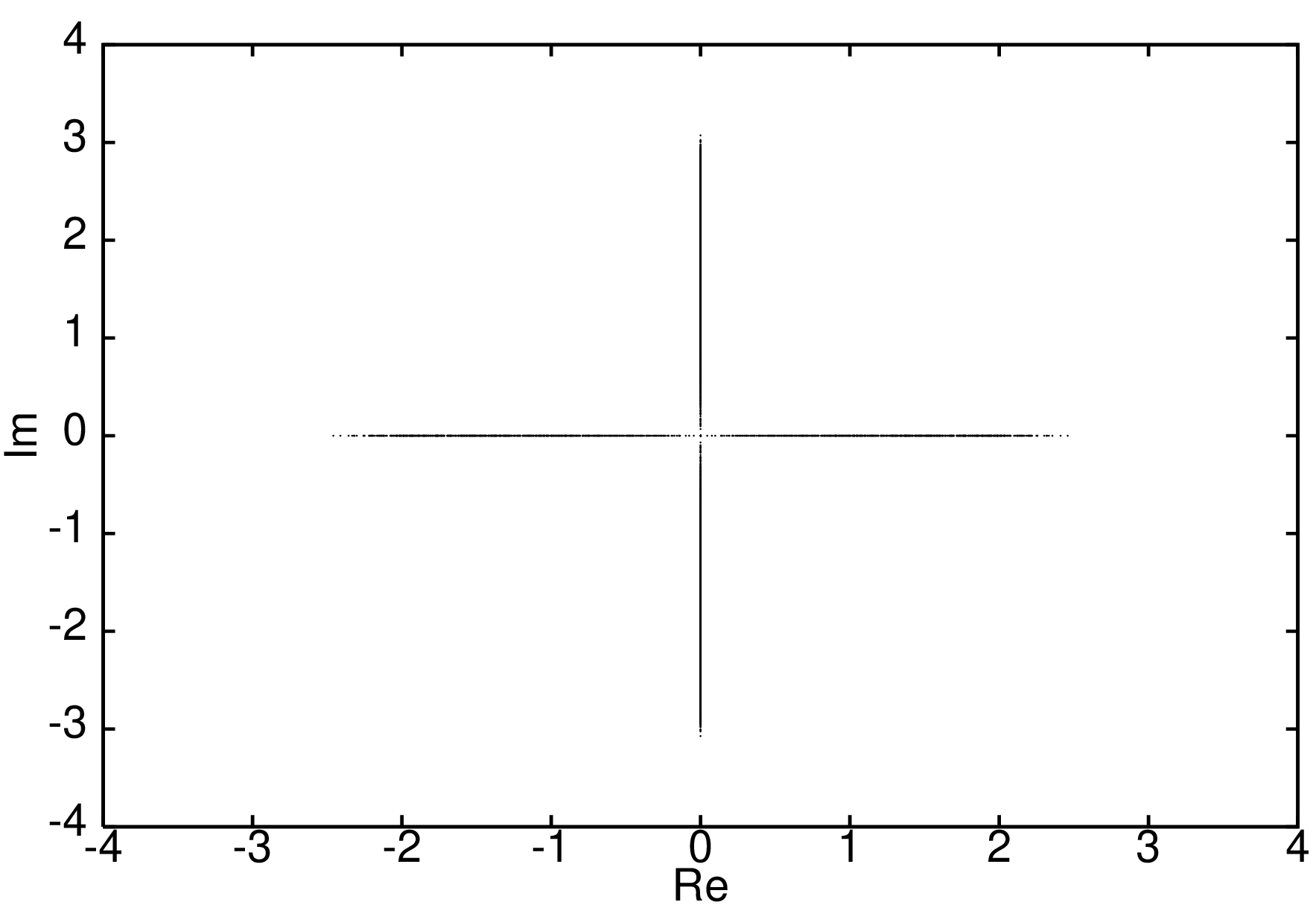,width=3cm,height=6.5cm}}\hspace{25mm}
{\psfig{figure=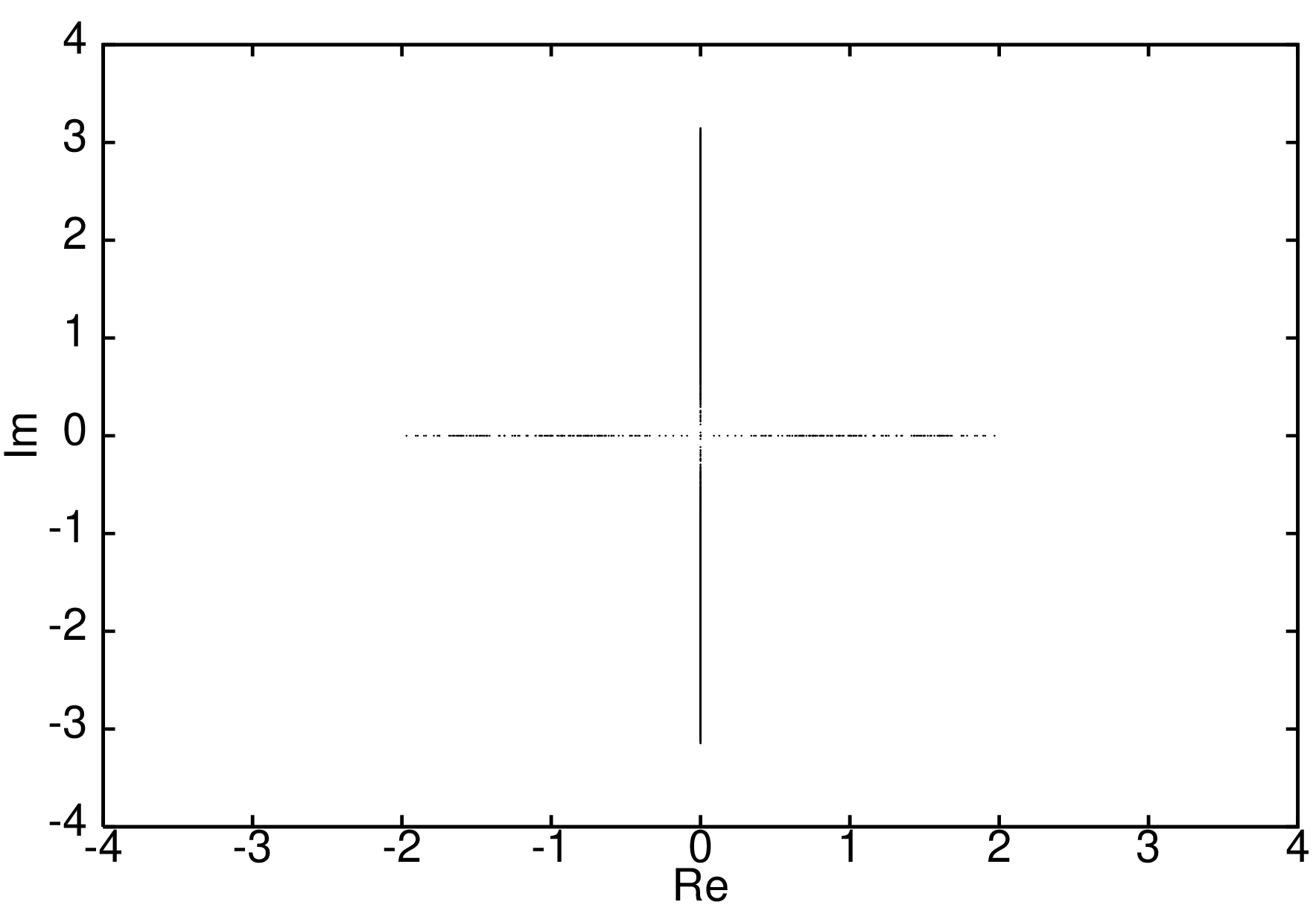,width=3cm,height=6.5cm}}}
\caption{Full complex eigenvalue spectrum of the monodromy matrix for $t=0$
         analyzing U(1) theory on a $4^3$ lattice 
         at $\beta=0.9$ in the confinement (left) and at $\beta=1.1$ in the
         Coulomb phase (right).
\label{Fig1}
 }
\centerline{{\hspace*{2mm}\psfig{figure=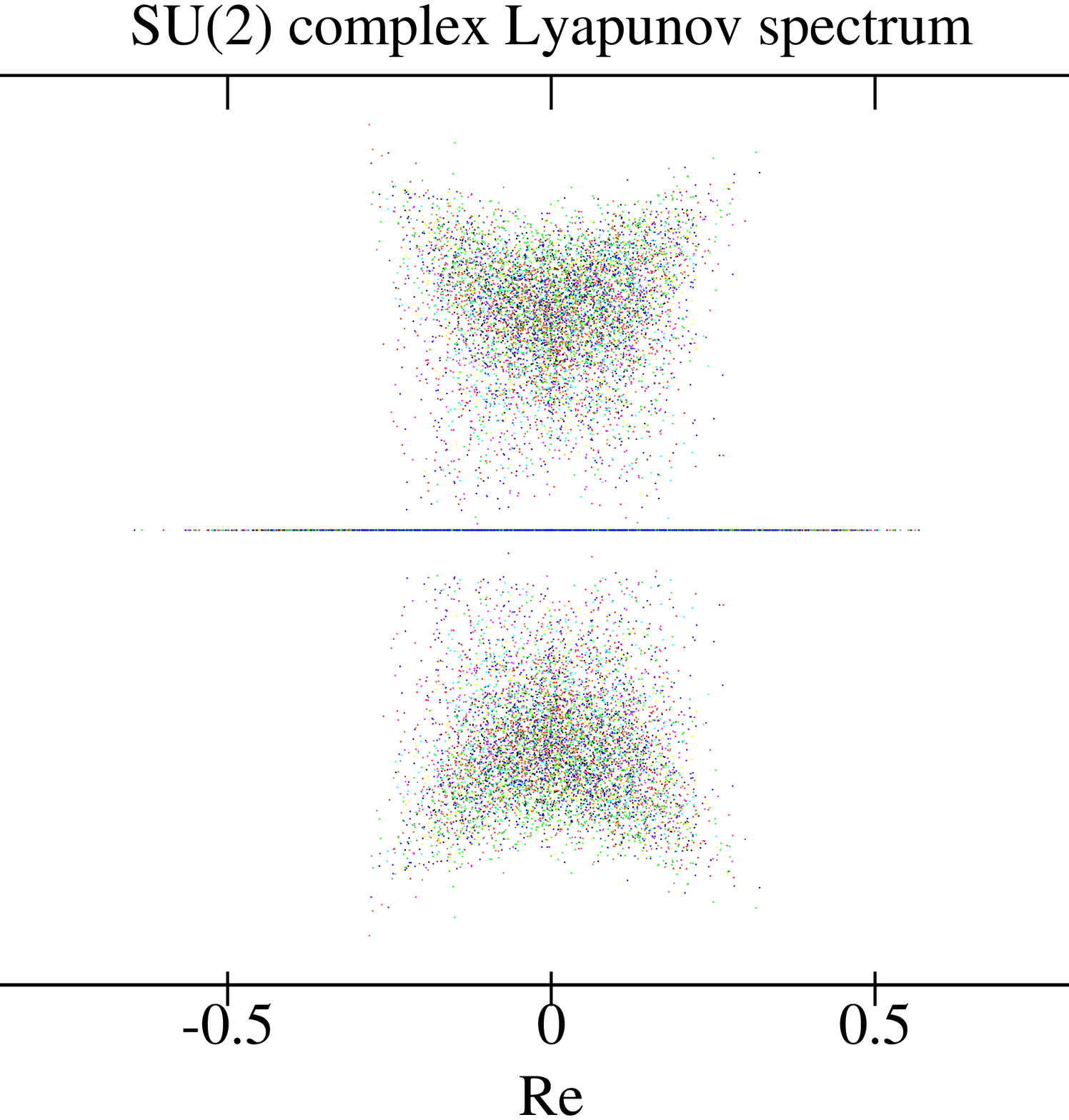,width=3cm,height=6.5cm}}\hspace{25mm}
{\psfig{figure=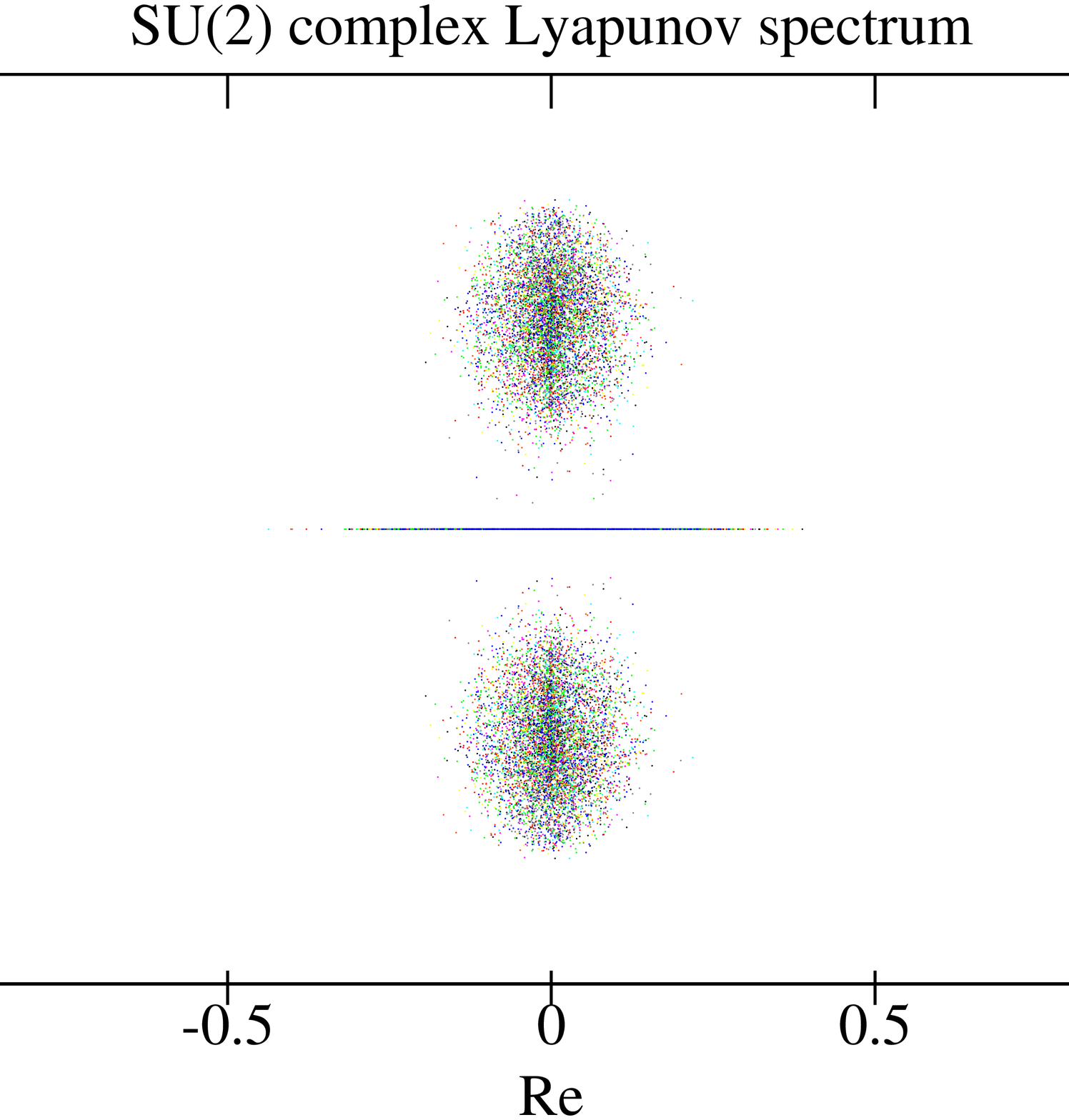,width=3cm,height=6.5cm}}}
\caption{Full complex eigenvalue spectrum of the monodromy matrix
         analyzing SU(2) theory on a $6^3$ lattice
         at high energy $g^2 aE=0.8$ (left) and at low energy $g^2 aE=0.1$
         (right)~\cite{FuBi01}.
\label{Fig2}
 }
\end{figure*}

Instead of the classical determination of the Lyapunov exponent
by the rescaling method outlined in the preceding
section, we now use the monodromy matrix
approach~\cite{FuBi01}. The Lyapunov spectrum $L_i$  is expressed in
 terms of the eigenvalues $\Lambda_i$ of the monodromy matrix $M$:
\begin{equation}
L_i = \lim_{t \rightarrow \infty}
\frac{1}{t}\int_0^t \Lambda_i(t')dt' \ , \;\;\;\;\;\;\; i=1,...,f \ ,
\label{LYAPU}
\end{equation}
where $\Lambda_i(t)$ are the solutions of the
characteristic equation for $f$ degrees of freedom
\begin{equation}
{\rm det} \left( \Lambda_i(t) 1 \!\;\!\! {\rm l} - M(t) \right) = 0
\label{CHAR}
\end{equation}
at a given time $t$.
Here $M$ is the linear stability matrix,
\begin{equation}
M = \left(\begin{array}{cc}
    \frac{\partial\dot{U}}{\partial U} &
    \frac{\partial\dot{U}}{\partial P} \\
 \\
    \frac{\partial\dot{P}}{\partial U} &
    \frac{\partial\dot{P}}{\partial P} \\ 
 \end{array} \right) \: = \: 
   \left(\begin{array}{cc}
    0 &
    1 \\
 \\
    -\frac{\partial^2S_3}{\partial U^2} &
    0 \\ 
 \end{array} \right)\: ,
\end{equation}
with the three-dimensional lattice action $S_3$.

\begin{figure}[ht]
\vspace*{6mm}
\centerline{{\hspace*{13mm}\psfig{figure=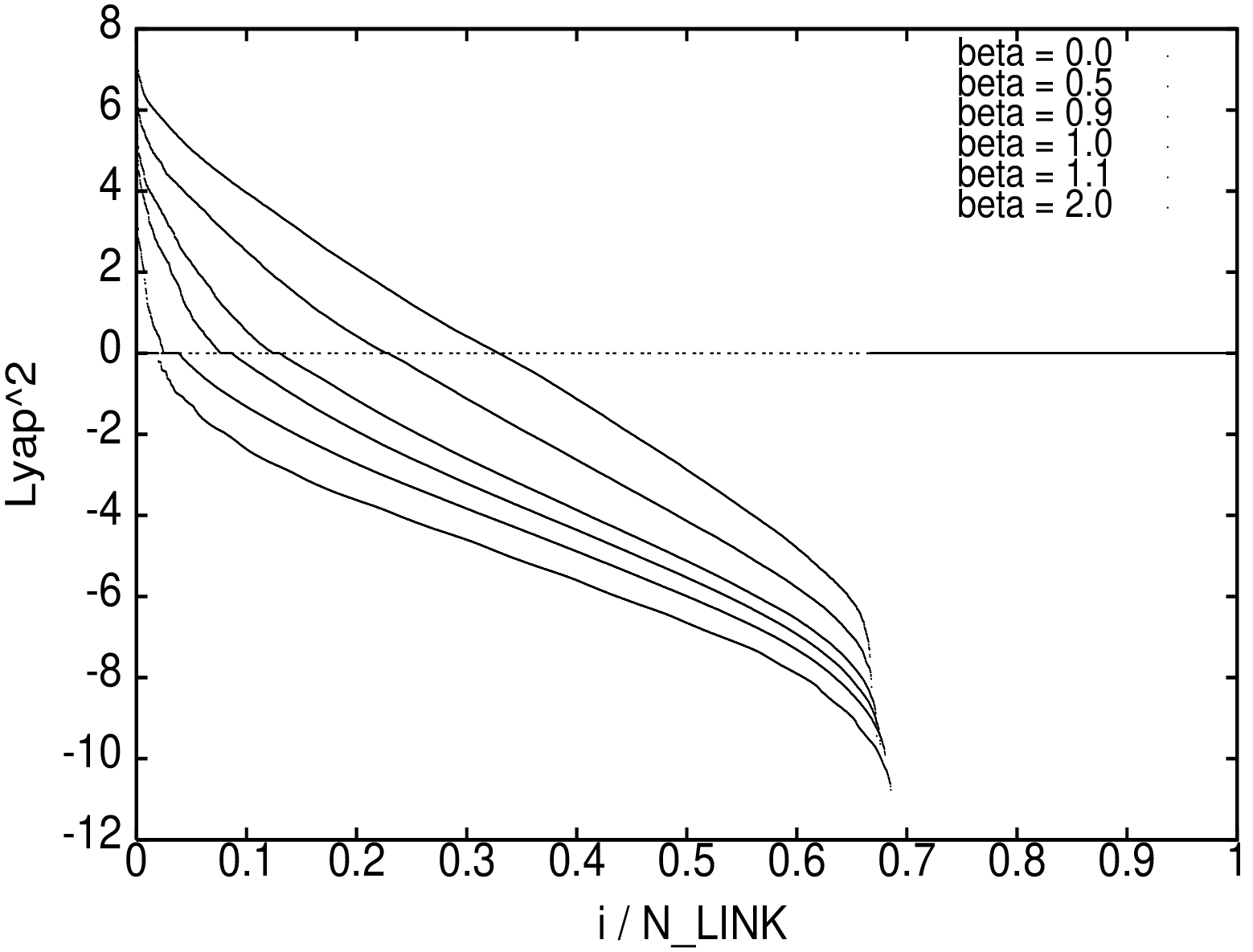,width=5.5cm,height=6.5cm,angle=-90}}
}\vspace*{-5mm}
\caption{Real part of the squared spectrum of the stability matrix decreasing
         with increasing $\beta$.
\label{Fig3}
 }
\centerline{{\hspace*{2mm}\psfig{figure=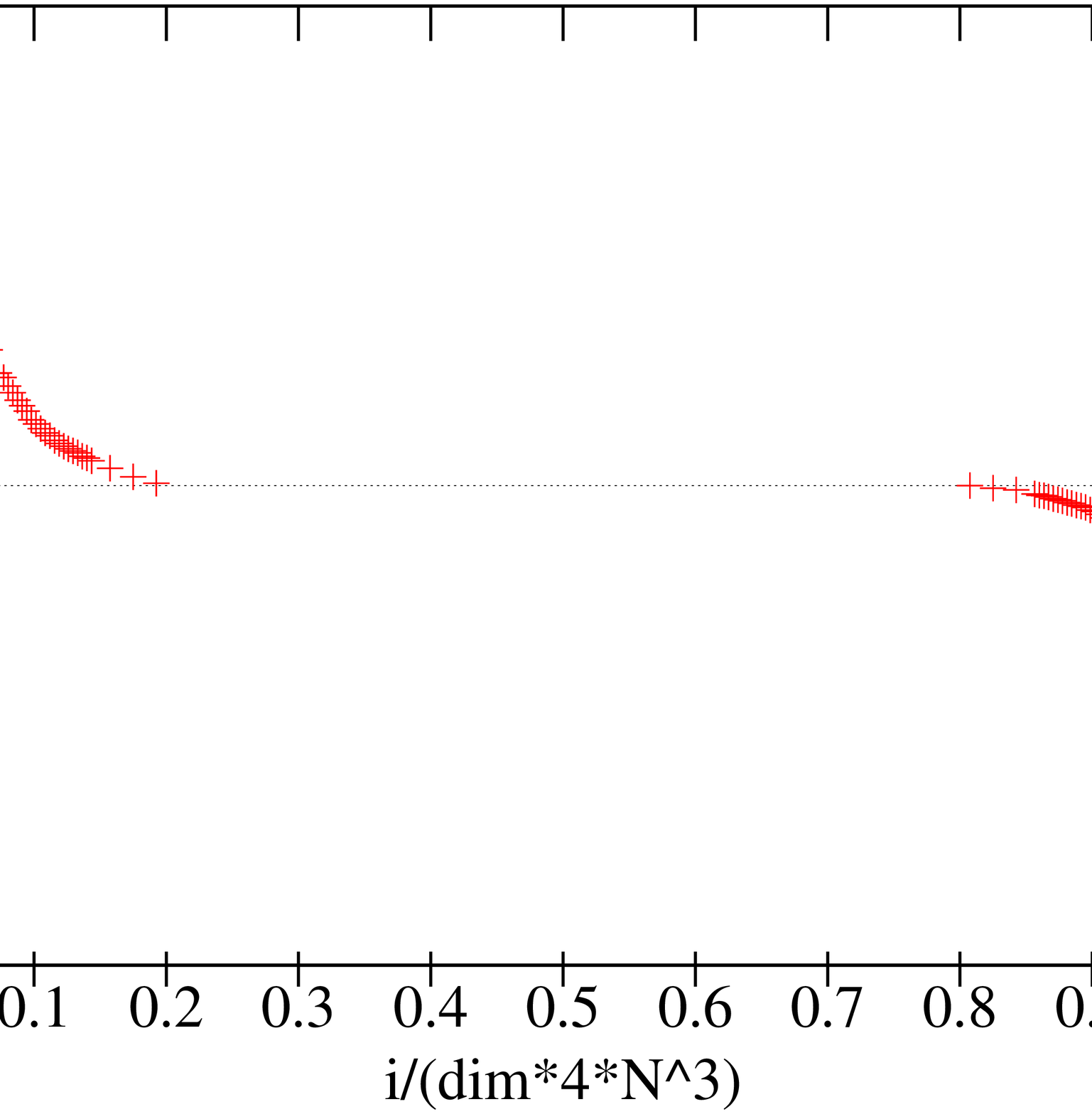,width=3.5cm,height=6.5cm}}
}
\caption{Real part of the spectrum of the stability matrix 
         with finite size scaling for increasing $N$~\cite{FuBi01}.
\label{Fig4}
 }
\end{figure}

Figure~\ref{Fig1} displays the complex eigenvalues for selected U(1)
gauge field configurations~\cite{MP2002}
prepared by a quantum Monte Carlo heat-bath algorithm.
By investigating the time evolution of the monodromy matrix it turned out
that the choice of the real time $t$ does not affect the eigenvalues appreciably.
In the confinement phase ($\beta=0.9$)
the eigenvalues lie on either the real or on the imaginary axes. This
is a nice illustration of a strange attractor of a chaotic system. Positive
Lyapunov exponents eject the trajectories from oscillating orbits provided by the
imaginary eigenvalues. Negative Lyapunov exponents attract the trajectories
keeping them confined in the basin.
In the Coulomb phase ($\beta=1.1$) the real Lyapunov exponents become rare to 
eventually vanish in the continuum limit~\cite{SCALING}.
In all cases the spectrum is symmetric with respect to
the real and imaginary axes: the former property
is due to the fact that the equations of motion are real, the latter
is due to the Hamiltonian being conservative (time independent).
Also a number of zero-frequency modes occur which is connected to
symmetry transformations commuting (in the Poisson bracket
sense) with the Hamiltonian, such as time independent gauge
transformations. 

Figure~\ref{Fig2} depicts the complex eigenvalues for SU(2)
gauge field configurations~\cite{FuBi01} taken from random
starts with a high energy $(g^2 aE=0.8)$ and a low energy
$(g^2 aE=0.1)$, respectively. Some caution is in order, since
the gauge fields were not obtained from a Monte Carlo equilibrium
configuration at definite coupling. Further, the entries of the
monodromy matrix were constructed from derivatives with respect
to components of the SU(2) matrix whereas the derivatives with
respect to the phase were taken in the U(1) case. The SU(2)
eigenvalues scatter in the complex plane but the symmetry of
the spectrum with respect to the real and imaginary axes is
still obvious. Again the prerequisites of a strange attractor
are fulfilled and one observes some tendency to regularity at
lower energy.

Figure~\ref{Fig3} shows the real part of the squared Lyapunov spectrum
of U(1) gauge fields for several couplings.
Since the U(1) spectrum is purely real or purely imaginary, its
square allows for a one-dimensional representation of its
distribution. 
It is symmetric for $\beta=0$ and is shifted to negative values towards
the Coulomb phase exhibiting an increasing number of zero modes with
increasing $\beta$.

Figure~\ref{Fig4} shows the real part of the Lyapunov spectrum
of SU(2) gauge fields~\cite{FuBi01} extrapolated to $1/N \to 0$ from data
taken at $N=2,3,4,5$ and $6$ at high energy $(g^2 aE=0.8)$.
Since the SU(2) spectrum lies inside the complex plane, here
only its real part is represented to gain a one-dimensional
distribution. The overall pattern resembles that obtained
earlier from smaller systems $(N=2,3)$ with the rescaling
method~\cite{Gong}.
The structure of the ordered real part of the Lyapunov spectrum
is similar at all energies considered, but the maximal point, $aL_{max}$,
scales with the energy, $g^2 aE$.

\section{Summary}

This contribution concentrated on the spectrum of the monodromy
matrix for (classical) compact U(1) theory.
It exhibited an interplay between positive, imaginary and negative
Lyapunov exponents in the confinement phase changing to a pure
imaginary spectrum deep in the Coulomb phase leading to the 
regularity of the Maxwell theory.
The spectrum for SU(2) theory was compared showing the
characteristics of a strange attractor. Since the configurations
were chosen from a random start corresponding to $\beta = 0$,
it will be interesting to compute the spectrum of the stability matrix
for finite $\beta$ and to investigate if the non-Abelian theory
stays chaotic in the continuum limit.


\begin{thebibliography}{10}

\bibitem{BOOK}  {\sc T.S. Bir\'o, S.G. Matinyan, and B. M\"uller:}
        {\em Chaos and Gauge Field Theory}, World Scientific,
        Singapore, 1995.
        %%CITATION = NONE;%%

\bibitem{SU2}  {\sc T.S. Bir\'o, M. Feurstein, and H. Markum},
        {\em Chaotic behavior of confining lattice gauge field configurations},
        APH Heavy Ion Physics 7 (1998), pp. 235-244;
        %%CITATION = HEP-LAT 9711002;%%
        {\sc T.S. Bir\'o, N. H\"ormann, H. Markum, and R. Pullirsch},
        {\em Chaos analyses in both phases of QED and QCD},
        Nucl. Phys. B (Proc. Suppl.) 86 (2000), pp. 403-407.

\bibitem{FuBi01} {\sc \'A. F\"ul\"op and T.S. Bir\'o},
                 {\em Towards the equation of state of classical SU(2) lattice gauge theory},
                 Phys. Rev. C64 (2001), 064902.
                 %%CITATION = HEP-PH 0107008;%%

\bibitem{MP2002} {\sc T.S. Bir\'o, H. Markum, R. Pullirsch, and W. Sakuler},
                 {\em Observables of lattice gauge theory in Minkowski space},
                 Nucl. Phys. B (Proc. Suppl.) 121 (2003), pp. 307-311.
                 %%CITATION = HEP-LAT 0210020;%%


\bibitem{SCALING}  {\sc L. Casetti, R. Gatto, and M. Pettini},
                   {\em Geometric approach to chaos in the classical dynamics of Abelian lattice gauge theory},
                   J. Phys. A32 (1999), pp. 3055-3067;
                   %%CITATION = CHAO-DYN 9810033;%%
                   {\sc H.B. Nielsen, H.H. Rugh, and S.E. Rugh},
                   {\em Chaos, scaling and existence of a continuum limit in classical non-Abelian lattice gauge theory},
                   hep-th/9611128;
                   %%CITATION = HEP-TH 9611128;%%
                   {\sc B. M\"uller},
                   {\em Study of chaos and scaling in classical SU(2) gauge theory},
                   chao-dyn/9607001;
                   %%CITATION = CHAO-DYN 9607001;%%
                   {\sc H.B. Nielsen, H.H. Rugh, and S.E. Rugh},
                   {\em Chaos and scaling in classical non-Abelian gauge fields},
                   chao-dyn/9605013.
                   %%CITATION = CHAO-DYN 9605013;%%

\bibitem{Gong} {\sc C. Gong},
               {\em Lyapunov spectra in SU(2) lattice gauge theory},
               Phys. Rev. D49 (1994), pp. 2642-2645.

\end{thebibliography}
\end{document}